\documentclass[sigconf,authorversion]{acmart}
\usepackage[utf8]{inputenc}
\pdfoutput=1
\usepackage{booktabs} 
\usepackage{fancyvrb}
\usepackage{pygsty}
\usepackage{natbib}
\usepackage{enumitem}

\setcopyright{acmlicensed}

\copyrightyear{2020}
\acmYear{2020}
\acmConference[CIKM '20] {The 29th ACM International Conference on Information and Knowledge Management}{October 19--23, 2020}{Virtual Event, Ireland}
\acmPrice{15.00}
\acmDOI{10.1145/3340531.3412778}
\acmISBN{978-1-4503-6859-9/20/10}

\begin{document}
\fancyhead{}

\title{LensKit for Python}
\thanks{The LensKit code, documentation, and community discussion forums are available from \url{https://lenskit.org}.}
\subtitle{Next-Generation Software for Recommender Systems Experiments}

\author{Michael D. Ekstrand}
\orcid{0000-0003-2467-0108}
\affiliation{%
  \institution{People \& Information Research Team\\ 
  Dept. of Computer Science,
  Boise State University}
  \city{Boise}
  \state{Idaho}
  \country{USA}
}
\email{michaelekstrand@boisestate.edu}

\begin{abstract}%
LensKit is an open-source toolkit for building, researching, and learning about recommender systems.
First released in 2010 as a Java framework, it has supported diverse published research, small-scale production deployments, and education in both MOOC and traditional classroom settings.
In this paper, I present the next generation of the LensKit project, re-envisioning the original tool's objectives as flexible Python package for supporting recommender systems research and development.
LensKit for Python (LKPY) enables researchers and students to build robust, flexible, and reproducible experiments that make use of the large and growing PyData and Scientific Python ecosystem, including scikit-learn, and TensorFlow.
To that end, it provides classical collaborative filtering implementations, recommender system evaluation metrics, data preparation routines, and tools for efficiently batch running recommendation algorithms, all usable in any combination with each other or with other Python software.

This paper describes the design goals, use cases, and capabilities of LKPY, contextualized in a reflection on the successes and failures of the original LensKit for Java software.
\end{abstract}

\ccsdesc[500]{Information systems~Recommender systems}
\ccsdesc[300]{Information systems~Evaluation of retrieval results}
\ccsdesc[300]{General and reference~Experimentation}

\keywords{recommender systems, evaluation, experiments, support software}

\maketitle

\section{Introduction}

The goal of the \textbf{LensKit project} is to develop and deploy tools and documentation that enable data processing, evaluation, and algorithms for recommender systems, with a particular focus on empowering research, education, and early-stage development.

For years, the original Java version of the LensKit toolkit \citep{ekstrandRethinkingRecommenderResearch2011}, the founding software of the LensKit project, has supported a wide array of research, educational, and deployment efforts.
Based on the experience my collaborators and I have using LensKit for research and teaching, in addition to feedback reports we received directly and indirectly from researchers and developers in industry and academia, we determined that the design and technology choices of the Java implementation were not a good match for the current and future needs of the recommender systems research community.

When we examined the software landscape to determine what existing tools might support the kinds of evaluation experiments people have carried out with LensKit and might run in coming years, it became clear to us that there  is a need for high-quality, well-tested support code for recommender systems experiments in a programming environment better suited for research and teaching.
Python data science libraries, including Pandas, scikit-learn, TensorFlow, and PyTorch, are widely used in recommender systems research and data science training, but there are not commonly-used implementations for experimental support code or evaluation metrics.
General-purpose machine learning tools, such as scikit-learn, often miss subtle recommender-specific evaluation details, such as splitting data by user instead of rating.
Existing recommender systems packages for Python (such as Surprise and PyRecLab) are often self-contained, like our Java LensKit code was, in that they drive the evaluation process and report a metric, instead of providing building blocks that the user can employ to build an experimental protocol that meets their precise needs.

To provide such tools in a maximally flexible fashion, I --- with support from my students and collaborators --- have developed LensKit for Python, a successor to the original Java code.
This new-generation toolkit brings the LensKit vision of \textit{reproducible research} supported by \textit{well-tested code} to a more \emph{widely-used} and \textit{easier-to-learn} computational environment.
LensKit code and examples are MIT-licensed, to minimize legal barriers to adoption.

In this paper, I begin by reflecting on the successes and failures of the LensKit project and its Java software to date.
I then present the goals, design, and use cases of the LensKit for Python (LKPY) software, along with a comparison to related tools.
I invite the community to provide feedback on how this software does or does not meet their needs and to participate in its ongoing development.

\section{Reflections on LensKit for Java}
\label{sec:reflect}

The Java version of LensKit saw active use from 2011 through 2018, supporting over 50 publications\footnote{A full list is at \url{https://lenskit.org/research}}, multiple production systems (including MovieLens \citep{harperMovieLensDatasetsHistory2015}), and the Recommender Systems MOOC on Coursera \citep{konstanTeachingRecommenderSystems2015}.
Reflection on my own personal experience and reports my collaborators and I received from members of the community yielded several insights into things that did and did not work for supporting researchers and educators.

\subsection{What We Got Right}

There are several things that I think we did well in the original LensKit software; some of these carry forward into the Python version, while others seem to have been good ideas in LensKit's original context but are not as important today.

\subsubsection{Testing}
In the LensKit development process, we have a strong focus on code testing \citep{ekstrandTestingRecommenders2016}.
This has served us well, and helped ensure the reliability of the LensKit code.
It is by no means perfect, and there have been bugs that slipped through, but the effort we spent on testing was a wise investment.

\subsubsection{The Java Platform}
When we began work on LensKit, there were three platforms we seriously considered: Java, Python, and C++.
At that time, the Python data science ecosystem was not what it is today; NumPy and SciPy existed, but Pandas was not yet well-known.
Pure Python does not have the performance needed for efficient recommender experiments.
Java enabled us to achieve strong computational performance in a widely-taught programming language with well-established standard practices, making it significantly easier for new users (particularly students) to adapt and contribute to it than I expect we would have seen with a corresponding C++ code base.
The earliest prototype code was in OCaml, but adoptability demanded a more widely-known language.

\subsubsection{Modular Algorithms}
LensKit was built around the idea of modular algorithms where individual components can be replaced and reconfigured.
In the item-item collaborative filter, for example, users can change the similarity function, the neighborhood weighting function, input rating vector normalizations, item neighborhood normalizations, and the strategy for mapping input data to rating vectors.
This configurability has been useful for exploring a range of configuration options for algorithms, at the expense of larger hyperparameter search spaces and increased software complexity.

\subsection{What Didn't Work}

Other aspects of LensKit for Java's design and development do not seem to have met our needs or those of the community as well.

\subsubsection{Opinionated Evaluation}
While LensKit's algorithms were highly configurable, its offline evaluation tools are much more opinionated.
Users can specify a few types of data splitting strategies, and recommendation candidate strategies, and it has a range of evaluation metrics, but the overall evaluation process and methods for aggregating metric results are fixed.
Metrics are also limited in their interface.

As my collaborators and I expanded our research into the recommender evaluation process itself, we repeatedly ran in to limits of this evaluation strategy. We often had to write new evaluation code in a fairly heavy framework or make the evaluator save intermediate files that we would reprocess in R or Python in order to carry out our research.
Too often the answer I would have to give to questions on the mailing list or StackOverflow is ``we're sorry, LensKit can't do that yet''.

One of our goals was to make it difficult to do an evaluation incorrectly: we wanted the defaults to embody best practices for offline evaluation.
However, best practices have been sufficiently unknown and fast-moving that we now believe this approach has held our research back more than it has helped the field.
It is particularly apparent that a different approach is necessary to support next-generation offline evaluation strategies, such as counterfactual evaluation \citep{bottouCounterfactualReasoningLearning2013}, and to carry out the evaluation research needed to advance understanding of effective, robust, and both internally- and ecologically-valid offline evaluations.

\subsubsection{Indirect Configuration}

LensKit for Java is built on the dependency injection principle, using a custom dependency injection container \citep{ekstrandDependencyInjectionStatic2016} to instantiate and connect recommender components.
This method gave us quite a few useful capabilities, such as automatically detecting components that could be shared between multiple experiment runs in a parameter tuning experiment.
The cost, however, was that it was difficult to configure an algorithm or to understand an algorithm's configuration.
The configuration visualization capabilities afforded by the DI framework were not enough to compensate for the complexity and subtlety of configuring an algorithm through dependency bindings.
It was also how to document how to use the system.

We believe this largely stems from the role of inversion of control in working with the LensKit for Java code --- users never write code that assembles an algorithm, they simply ask LensKit to instantiate one and LensKit calls their custom components.
It is therefore difficult to develop an intuition about how LensKit works and how to realize a desired configuration.

\subsubsection{Implicit Features}
Beyond indirect configuration, LensKit has a lot of implicit behavior in its algorithms and evaluator.
This has at least two downsides: first, it is less clear from reading a configuration precisely what LensKit will do, making it more difficult to review code and experiment scripts; second, if documentation slipped behind the code, understanding the behavior of LensKit experiment scripts required reading the LensKit source code itself.

\subsubsection{Living in an Island}
LensKit has its own data structures and data access paradigms.
Part of this is due to lack of standardized, high-quality scientific data tooling that is not connected to a larger framework such as Spark (while Spark does seem to expose data structures as a separate library, they are not well-documented).

This made it difficult, however, to make LensKit interoperate with other software and data sets.
While LensKit for Java is flexible in the data it accepts, users must write data adapters, and common data layers such as Hibernate had unacceptable performance penalties for research experiments.
Integrating with other tools such as Spark is difficult.
It's unclear what, given the Java commitment, we could have done differently here, but it is definitely a liability for the future of recommender systems research.

\section{Design Goals}
Based on my experience with the Java software and sense of the current landscape of recommender systems research software, I identified several design goals to make LKPY as useful and usable as possible.
These goals include:

\begin{description}[leftmargin=1em]
\item[Build on Standard Tools]
There are now standard tools, such as Pandas and the surrounding PyData ecosystem \citep{mckinneyPythonDataAnalysis2018}, that are widely adopted for data science and machine learning research.
Using them maximizes interoperability with other software packages and enables code reuse across different types of research.
Jupyter notebooks are also a valuable means of promoting reproducibility, and my team uses experiment workflows where the final analysis consists of loading recommender outputs into a Jupyter notebook and computing desired metrics over them.
Small experiments can even be driven entirely from Jupyter.

\item[Leverage Existing Software]
Modern recommender systems are usually machine learning systems; a recommender research tool\-kit should not try to reinvent that wheel.
Scikit-Learn provides many machine learning algorithms suitable for recommender systems research, and PyTorch and TensorFlow provide general-purpose optimization for modular differentiable machine learning models.
Recommender research tooling should work seamlessly with algorithms implemented in these kinds of toolkits.

\item[Expose the Data Pipeline]
Many recommendation packages, including the Java version of LensKit, hide or at least take ownership of the data pipeline: they control data splitting, algorithm training, recommendation, and evaluation.
Outputs of each stage can be examined, but not manipulated, and the pipeline itself cannot be easily changed.
Putting the user in control of the pipeline, and providing functions to implement standard versions of each of its stages, has two benefits: the actual pipeline used is clearly documented in the experiment code, and the pipeline can be modified as research needs demand.

\item[Explicit is Better than Implicit]
Python has a long-standing philosophy of explicitly denoting desired operations instead of depending on implicitly-defined behavior.
Applying this to research code would, I expect, make it easier to review experiment designs and improve the reliability and rigor of reproducible research.

I also posit that exposing the data pipeline will also make the LKPY code itself easier to read and understand.

\item[Simple Interfaces]
Interfaces to individual software components should be as simple as possible, so that it is easy to document, test, and reimplement them.
They should also be data-agnostic, when feasible, to avoid unnecessary limits the types of data or algorithms that can be used with LKPY.

\item[Ease of Development]
For long-term viability, the development process, not just the resulting software, needs to approachable for prospective users and contributors.
I particularly want to ensure that students are able to contribute to LensKit so it can benefit directly from their research.
Therefore, the development tools should be as standard and easy-to-use as possible.
\end{description}

Notably absent from this list is the Java version of LensKit's (in)famous algorithm configurability.
That configurability was useful for exploring the space of algorithm configurations, but its particular design is more suitable to heuristic techniques such as k-NN; machine learning approaches seem better served by a different design.
Connecting with existing flexible optimization software such as TensorFlow and PyTorch will provide a great deal of configurability for new algorithms.
I think it is more important for recommender-specific software to focus on flexibility in the experiment design to facilitate education and new types of research.

\section{LensKit Use Cases}

LensKit for Python directly supports several use cases, and has a modular and loosely-coupled design to support many more uses cases in the future.
In this section I describe use cases explicitly supported by LensKit for Python, along with available public work that uses them, as well as some use cases that we are working on supporting in the very near future. 

\subsection{Recommender System Evaluation}
LensKit enables offline evaluation of the effectiveness of new and improved algorithms through several features:

\begin{itemize}
    \item Support for bulk-running recommendations and predictions for comparison with test data.
    \item Evaluation code to compute common prediction accuracy and top-$N$ accuracy metrics.
    \item High-quality, well-tested implementations of classic collaborative filters for use as baselines.
    \item Easy-to-implement interfaces to enable head-to-head comparisons between new approaches and baseline or prior state-of-the-art techniques.
    \item Easily usable in and with other tools, such as Jupyter for reporting experiment results and scikit-optimize for hyperparameter search.
\end{itemize}

These features are loosely coupled, so they can be recombined, adapted, and individually replaced to meet the needs of a wide range of research designs.
Unlike the Java version of LensKit and many other recommendation tools, the structure of an evaluation and the relationship of data to metrics is not prescribed by LKPY.

LensKit for Python has seen published use in this capacity to study recommendation techniques for children's books \citep{Ng2020-gf}, in addition to our demonstration experiment that compares the accuracy of different recommender algorithms on a several public data sets\footnote{\url{https://github.com/lenskit/lk-demo-experiment}}.

\subsection{Offline Recommender System Research}
The same facilities that make LKPY useful for evaluating new recommendation techniques are also valuable for other experiments that study the behavior of recommender algorithms on available data, the characteristics of experimental protocols, and many other research objectives around recommender systems and their data.

My own research team has used LKPY to study errors in evaluation protocols \citep{Tian2020-vs}, and to rebuild the experiments from our work on author gender biases \citep{Ekstrand2018-um} for an expanded version currently under review.
\citet{Narayan2019-ge} used LKPY to study the effect of rating obscure items, and we expect more such projects to use the software in the coming years.

\subsection{Education}
LensKit is conceived as a toolkit that is meant to be useful in educational settings, as an environment for teaching recommender systems and personalization at either the undergraduate or graduate levels.  I have used it successfully to support a graduate-level class (Boise State CS 538 --- \emph{Recommender Systems and Online Personalization}), and have received reports of others successfully using it as well.

LensKit is also suitable for online demonstrations.
PBS Digital Studios' \emph{Crash Course: Artificial Intelligence} program used it to demonstrate recommender systems in their video and accompanying notebook on Google Collaboratory \citep{PBS2019-lr}.

\subsection{Online Studies (In Progress)}
\label{subsec:OnlineStudies}
We want to be able to use LensKit for Python in online user studies as well as offline experiments. 

Right now, online experiments are possible  by embedding the Python code in a web application built with a framework like Flask or Django.
However, this does not scale well and requires tedious and error-prone code.
To address this limitation, my student and I are currently developing a recommendation server that exposes LKPY's recommender APIs as a web service.
When completed (expected by late 2020), using any LKPY algorithm --- or any other algorithm that implements the required Python interfaces --- will be as easy as making a REST call.

\subsection{Small-Scale Production (In Progress)}
The same recommender server that will enable online studies (see Section \ref{subsec:OnlineStudies}) will also make it easy to use LensKit for Python in small- to medium-scale production services and prototypes.
Large-scale services will likely require other facilities, but startups and research groups will be able to use LKPY to prototype recommendation ideas and move to other software as their needs grow.

\section{LensKit Software Description}

To support these design goals and use cases, LKPY provides:

\begin{itemize}
    \item Data preparation for recommender experiments
    \item Common APIs for recommendation algorithms
    \item Batch-processing functions for carrying out experiments
    \item Evaluation metrics
    \item A suite of collaborative filtering algorithms
\end{itemize}

\subsection{Example Code}

\begin{figure}[tb]
\input{example-py}
\caption{Example of a simple evaluation.}
\label{fig:example}
\end{figure}

Figure \ref{fig:example} shows an example of using LKPY to compute the normalized discounted cumulative gain of a matrix factorization algorithm with 5-fold cross-validation on the MovieLens 100K data set \citep{harperMovieLensDatasetsHistory2015}, consisting of 100K movie ratings from just under 1000 users.
It is somewhat verbose, but every step of the evaluation process is clear in the code.
This allows the experiment structure to be reviewed during the research process and makes the code self-documenting when it is published with the final paper.
Further, since the data at each stage is in standard Pandas and NumPy data structures, it can be inspected, analyzed, and transformed at any point to support arbitrary research objectives.

\subsection{Core Interfaces and Data Expectations}
LKPY's algorithm-related code uses the \texttt{Algorithm} interface, based on the patterns used by Scikit-Learn~\citep{buitinckAPIDesignMachine2013}, where a \texttt{fit} method estimates model parameters (trains the model) in-place.
This makes the API familiar to users with experience using other packages in the SciKit ecosystem.

Common recommendation operations are defined by additional interfaces that subclass \texttt{Algorithm}:

\begin{description}[leftmargin=1em]
\item[Predictor] implements user-item preference prediction; implementations using explicit-feedback data typically return predicted ratings, while other algorithms return purchase probabilities or other scores useful for ranking results.
The \texttt{predict\_for\_user} method takes a user ID and a list of items and returns predictions for the specified items. 
It can optionally take a Pandas series of user ratings; some algorithms are able to use these ratings to provide recommendations that are responsive to user ratings that were not available when the model was trained.

\item[Recommender] implements top-$N$ recommendation. 
It defines the \texttt{recommend} method, which takes a user ID and optionally a desired list length and/or list of candidate item IDs and returns a ranked recommendation list, along with scores if applicable.
It also can accept a series of ratings.

\item[CandidateSelector] is usually not used alone, but selects candidate items for recommendation when no candidate set has been provided to \texttt{recommend}.
\end{description}

The \texttt{fit} method takes a \texttt{ratings} argument containing user-item preference data.
This argument is a Pandas \texttt{DataFrame} with ``user'' and ``item'' columns; explicit-feedback rating data stores the rating value in a ``rating`` column.
LensKit code --- including both algorithm implementations and infrastructural code that passes data to them --- makes no assumptions about the (non-)existence of other columns, and also allows additional named arguments to \texttt{fit} that the built-in algorithm implementations ignore, to allow arbitrary additional data to be passed to algorithm implementations.
Therefore, while LensKit's current algorithm implementations are focused on collaborative filtering, content-based, context-aware, and other types of recommendation techniques can be readily tested.  For example, a review-based recommender can require \texttt{ratings} to also have a ``review'' column, or a content-based recommender can take an additional \texttt{item\_data} parameter with item side information; LensKit code in the call chain between the user's experimental code and their algorithm's \texttt{fit} method will ignore and pass on data it doesn't recognize.

A large number of recommendation algorithms, even many learning-to-rank techniques, compute user-personaized scores for individual items and then rank items by score for the final selection (e.g.\ even Bayesian Personalized Ranking \citep{rendleBPRBayesianPersonalized2009} learns a function $s(i|u)$ by optimizing it to correctly rank items).
Implementers of such algorithms will typically implement the \texttt{Predictor} interface, and use it with LensKit's top-$N$ wrapper to produce item recommendation lists.
Algorithms that directly produce ranked lists, or perturb existing ranked lists (such as diversification strategies \citep{Ziegler2005-zo}), will implement the \texttt{Recommender} interface.
This separation enables code reuse (top-$N$ logic is the same no matter how the underlying scores are computed), and allows researchers to easily and independently experiment with new ideas for either item scoring or aggregating scored items into recommendation lists.

The LKPY interfaces are designed to support both batch operation --- LensKit's current primary usage mode in an offline experiment --- and per-user recommendation for online settings.
As I discuss in \S\ref{subsec:OnlineStudies}, the online use case is not yet as easy to deploy as we would like, but LKPY's algorithms and new algorithms others develop with its interfaces will be immediately usable once we complete the required infrastructural development (\S\ref{sec:future}).

\subsection{Experiment and Evaluation Support}
As we illustrate in Figure~\ref{fig:example}, a typical offline recommender system experiment consists of three major steps:

\begin{enumerate}
    \item Prepare training and test partitions of user ratings, interaction traces, or other recommender input data.
    \item Train algorithms and produce recommendations or predictions for test users.
    \item Compute metrics over the algorithm output
\end{enumerate}

LKPY provides code for each of these steps.

The \texttt{crossfold} module provides mechanisms for preparing multiple train-test partitions of ratings or similarly-structured data.
Just partitioning the ratings, as a standard machine learning cross-validation package would do, does not result in an accurate simulation of the actual application goals of a typical recommender system \citep{Gunawardana2009-qy}.
LKPY's crossfolding support focuses on test \emph{users}, either partitioning the users (as in cross-validation) or producing disjoint sets of test users (for when evaluating on all users is cost-prohibitive).
For each test user, it will select a set of test ratings (either randomly or using timestamps) and place the user's other ratings in the training set, along with the ratings from users not in the test partition, so the model can learn the test user's preferences.
We also provide functions to partition or sample ratings, and to perform item-based sampling as required by certain protocols \citep{Bellogin2012-yf}.

The \texttt{batch} module provides functions to mass-produce predictions or recommendations, with support for parallelizing over test users, to facilitate the second step of the experiment.
It provides both batch prediction, where the input is an algorithm and a data frame of (user,item) pairs (if the data frame also contains ratings, those are included in the prediction output to ease prediction accuracy calculation), and batch recommendation using an algorithm and a list of user IDs.
The output of both is a Pandas data frame that can be analyzed directly or serialized for later use.

Finally, the \texttt{metrics} and \texttt{topn} modules provide evaluation support.
Prediction accuracy is relatively straightforward to support: after merging the prediction data with the rating data, MAE, RMSE, and other metrics can be computed with Pandas aggregate operations.
The LKPY's \texttt{rmse} and \texttt{mae} functions are usable in this setting, and can be used to compute either per-system or per-user accuracy.

Top-$N$ recommendation list accuracy is more difficult to implement correctly in the general case.
Precision, hit rate, average precision, and reciprocal rank can be computed by taking a left join between the recommendation list and the test ratings and then computing the metric; other metrics, such as normalized discounted cumulative gain and recall, require access not only to the recommended items and their relevance information, but also to the test items that were \emph{not} recommended.
Top-$N$ evaluation metric are implemented as functions of two parameters --- a recommendation list and the user's test ratings or other relevance data --- and return a metric value, so users can implement new metrics easily.

The \texttt{RecListAnalysis} class groups recommendation results by user, algorithm, data set, and other grouping variables, matches them with test ratings, and provides both recommendation lists and test ratings to the metric to facilitate correct computation.

\subsection{Algorithm Implementations}

Non-personalized algorithms are useful as naive baselines, to establish bounds on the performance of more sophisticated algorithms, and as fallbacks or building blocks for other techniques. Non-personalized algorithms in LKPY include:

\begin{description}[leftmargin=1em]
\item[Popular] recommends the most frequently-consumed items a user has not interacted with.
\item[Bias] implements the user-item bias rating predictor \citep{Ekstrand2010-wg}.
\item[Random] recommends randomly-selected items.
\end{description}

The $k$-NN implementations provide classic neighborhood-based collaborative filtering, and work with both explicit ratings and implicit-feedback consumption data:

\begin{description}[leftmargin=1em]
\item[UserUser] implements the user-user algorithm \citep{herlockerEmpiricalAnalysisDesign2002}, using cosine similarity over mean-centered rating vectors.
The similarity function is fixed to facilitate performance and because we have found it to work remarkably well over a wide range of data sets \citep{ekstrandRethinkingRecommenderResearch2011}.

\item[ItemItem] implements the item-item algorithm \citep{sarwarItembasedCollaborativeFiltering2001, deshpandeItembasedTopNRecommendation2004}, also with cosine similarity over mean-centered vectors.
\end{description}

Both $k$-NN algorithms have configurable neighborhood sizes and rating aggregation strategies.

LensKit also provides matrix factorization implementations for both explicit and implicit feedback:

\begin{description}[leftmargin=1em]
\item[BiasedMF] provides biased matrix factorization for explicit feedback learned with alternating least squares \citep{zhouLargeScaleParallelCollaborative2008}, optimized with coordinate descent \citep{Pilaszy2010-ir}.
\item[ImplicitMF] provides ALS matrix factorization for implicit feedback data \citep{huCollaborativeFilteringImplicit2008} using the conjugate gradient method \citep{Takacs2011-ix}.
\item[FunkSVD] implements Simon Funk's gradient descent biased matrix factorization \citep{funkNetflixUpdateTry2006}.
\item[svd] provides biased matrix factorization using the truncated SVD algorithm from SciKit-Learn \citep{scikit-learn}.
\item[tf] implements various matrix factorization models using TensorFlow \citep{tensorflow2015-whitepaper}. This includes biased MF for rating prediction and Bayesian Personalized Ranking (BPR,  \citep{rendleBPRBayesianPersonalized2009}) for learning-to-rank. Biased MF comes in two versions: one that fits matrices to the residuals of the \texttt{Bias} model, and another that jointly learns biases and latent feature vectors.
\end{description}

The TensorFlow and Scikit-Learn implementations serve both as baselines for evaluating new algorithms and as examples for developing new proposed recommendation techniques with these libraries within the LensKit framework.

Finally, LKPY provides implementations of its algorithm APIs using Implicit\footnote{\url{https://github.com/benfred/implicit}} (Implicit ALS and BPR with GPU acceleration) and HPFREC\footnote{\url{https://github.com/david-cortes/hpfrec}} (providing Hierarchical Poisson Factorization~\citep{gopalanScalableRecommendationPoisson2013}).

The LensKit project's primary goal is to make it easy to carry out experiments with algorithms from anywhere, so new algorithms are not our first priority; we do, however, welcome both new implementations and bridges to other packages.

\subsection{Dependencies and Environment}

LensKit for Python leverages Pandas \citep{mckinneyPythonDataAnalysis2018}, numpy \citep{oliphantGuideNumPy2006}, and scipy \citep{oliphantPythonScientificComputing2007}, along with several other Python modules.
We use Numba \citep{lamNumbaLLVMbasedPython2015} for native-code acceleration in custom algorithm implementations, as it enables efficient, parallelized NumPy-based code without complex build toolchains.
We plan to make more use of TensorFlow for the optimization process of future algorithm implementations.
The batch recommendation and predictions routines use Python's multi-process concurrency support, combined with the shared-memory facilities in the \texttt{binpickle} package, to parallelize over users when generating recommender output for an offline experiment.

LKPY supports recent versions of Python 3 on Windows, Linux, and macOS.
Project policy is to support the oldest version of Python 3 available in the most most recent stable release of major Linux distributions (Ubuntu LTS, RHEL/CentOS, and Debian).
We optimize for Anaconda-based Python installations (including direct use of certain Intel MKL capabilities, with pure-Python SciPy fallbacks when MKL is not available), but also support vanilla Python on Linux and Windows\footnote{We only support Anaconda-based Python installs on macOS at this time due to the difficulty of supporting scientific packages on system Python.}.
We provide binaries via Anaconda Cloud\footnote{\url{https://anaconda.org/lenskit/lenskit}} and source distributions on the Python Package Index\footnote{\url{https://pypi.org/project/lenskit/}}.

\subsection{Use Across Scales}
LensKit supports both research and educational use at multiple scales.
Simple experiments and demos can be done entirely in a single Jupyter or Collaboratory notebook, as shown by PBS \emph{Crash Course} and the Quick Start guide in the project documentation.

As a project scales up, the recommendation code can move into scripts that produce results for subsequent analysis.
Data pre-processing, model training, output generation, output aggregation, and final performance evaluation can all be split into separate stages with output written to intermediate files.
This allows, for example, research teams to use a compute cluster to train models and generate recommendations, and download the resulting files to a local workstation for computing evaluation metrics and producing the final analysis report.

LKPY's algorithm implementations are focused on single-node performance, and can scale with data that can be processed on a single node (which can be remarkably large when sufficent memory is available).
For even larger data sets, users can develop their recommendation techniques in the distributed processing framework of their choice, such as Dask, Spark, or TensorFlow, and use LensKit's data processing and/or evaluation techniques to prepare and analyze the experiment and its results.
LKPY aims to grow with its users, and to progressively get out of their way as they outgrow its built-in capabilities.

\subsection{Quality Assurance}
LensKit for Python includes an extensive test suite, following the testing practices that we found successful with the Java software \citep{ekstrandTestingRecommenders2016}.
These tests include:

\begin{itemize}
    \item Unit tests for individual functions and modules.
    \item End-to-end model build and serialization tests for all algorithm implementations.
    These tests train the model on the MovieLens Latest Small data set \citep{harperMovieLensDatasetsHistory2015}, a snapshot of which is included in the LensKit source tree, and confirm that the algorithm can successfully build, be saved to disk, and reloaded into memory without crashing or exhibiting other unexpected behavior.
    \item Accuracy tests that cross-validate models on the ML-100K data set and check that their rating prediction and top-$N$ recommendation accuracy is within the expected range for that algorithm on that data set.
    \item Known-output tests that train the algorithm on a small MovieLens data set and compare its output with known-correct output.  This is particularly for algorithms expected to exhibit the same behavior as their Java predecessors.
\end{itemize}

GitHub Actions run the test suite on all supported OS families and Python versions and measure test coverage.
New features and fixes are implemented in feature branches and merged into the main development tree through pull requests, ensuring tests are run on all new code prior to integration.
CodeClimate lint helps maintain code standards and catch some potential errors.

\subsection{Documentation and Examples}
LensKit for Python's tutorial and reference documention includes:

\begin{itemize}
    \item A Getting Started guide that implements a simple recommender accuracy experiment in a self-contained Jupyter notebook, available on Google Collab.
    \item A user's guide documenting LKPY's features APIs.
    \item An example experiment, published on GitHub, that demonstrates a more realistic comparison of the effectiveness of recommendation algorithms on public data sets.
    \item Source code for experiments using LKPY \citep{Tian2020-vs, Ekstrand2019-iu}.
\end{itemize}

The examples also demonstrate LensKit use across a range of scales, from a single notebook to a full experiment that saves trained models and intermediate data to disk and coordinates experimental runs with Data Version Control \citep{Kuprieiev2020-xh}.
The experiment for our extended author gender paper \citep{Ekstrand2019-iu} is completely reproducible end-to-end with a single command (\texttt{dvc repro}) and approximately 2 days of compute time.
It also demonstrates automatic hyperparameter optimization with scikit-optimize \citep{tim_head_2018_1207017}.

\section{Future Directions}
\label{sec:future}

LensKit for Python development is ongoing, and there is much to add to the software.
Two immediate priorities are to:

\begin{itemize}
\item Complete the recommendation web service to support online experiments and deployed applications.

\item Build example content- and review-based recommendation algorithms to demonstrate working with additional data sources in the LensKit environment.
\end{itemize}

While LKPY develompent to date has primarily been by myself with support from the People and Information Research Team and collaborators, we welcome contributions from anyone subject to the project's licensing requirements (MIT license) and code quality standards.
Development and discussion are open to the public on GitHub and Google Groups.

\section{Comparison to Existing Packages}

Before building LKPY, I re-examined the landscape of existing software to determine if the needs we saw would be met by one of the other software packages.
I was unable to find existing tooling that supports the LensKit project's goals of flexible recommender systems experiments that leverage the PyData ecosystem.

Two of the most closely-related Python toolkits are surprise \citep{hugSurprisePythonLibrary2017} and PyRecLab \citep{sepulvedaPyRecLabSoftwareLibrary2017}.
Surprise uses numpy and scipy, but not the fuller PyData ecosystem (such as Pandas) and has its own data set management.
PyRecLab has extensive algorithm support and is an effective toolkit for teaching and rapid prototyping, but it is a C++ library with a Python API, rather than a native Python toolkit of reconfigurable pieces, so it doesn't use PyData tools or integrate with packages like TensorFlow, making it less appropriate for supporting the kinds of research I want to enable.
Our research requires us to be able to write out the evaluation steps in our own code so that we can experiment with them more readily, and implement new algorithms without developing C++ proficiency.

The LKPY approach is perhaps most similar to that of mrec\footnote{\url{https://github.com/Mendeley/mrec}}, focusing on discrete steps enabled by separate tools.
mrec, however, does not leverage contemporary Python data science tools, and does not seem to be under active maintenance.
LKPY also focuses on Python as the scripting language for experiment control while mrec favors command-line tools orchestrated with shell scripts.

Many Python-based recommender systems research project seem to implement their their own evaluation procedures directly while building the recommendation algorithm in scikit-learn or one of the deep learning frameworks.
LKPY integrates with such workflows, enabling them to leverage common, well-tested implementations of metrics and other experimental support code while continuing to use their existing data flows for the recommendation process.

\section{Conclusion}

My collaborators and I have learned many lessons developing, maintaining, using, and teaching the Java-based LensKit software.
We came to the conclusion that, in its current form, it is not the needs of the recommender systems community well, and resources would be better spent on improving the research being done with the widely-used Python packages that drive much of modern data science.
LensKit for Python (LKPY) provides code to make it easy for researchers, students, and developers to build and run experiments on recommender systems and prototype new ideas while leveraging the extensive software, documentation, and tutorials available for data science and machine learning in Python.

Where Java LensKit focused on providing building blocks for recommender \emph{algorithms}, LensKit for Python provides building blocks for recommender \emph{experiments}.
Built on the PyData stack and oragnized around clear, explicit data processing pipelines with a minimum of bespoke concepts, I believe LKPY provides a solid foundation for new experimentation and concepts at all stages of the recommender systems research lifecycle.
It is already seeing adoption in both research \citep{Wang2019-ed, Narayan2019-ge, Balseca_Ninez2019-yy} and instructional settings \citep{Varga2019-vs, PBS2019-lr}, in addition to supporting my own group's research and educational efforts.
We welcome contributions on GitHub, and invite feedback from the community to guide future development.

\begin{acks}
This material is based upon work supported by the National Science Foundation under Grant No. IIS 17-51278.
I thank my collaborators in the People and Information Research Team (PIReT) for their support in this project, particularly Sole Pera for her comments on the manuscript and my students Amifa Raj, Mucun Tian, and Carlos Segura Cerna for pushing the software forward; the GroupLens Center for Social Computing, particularly collaborators John Riedl \& Michael Ludwig for supporting the Java software and Joe Konstan \& Daniel Kluver for supporting both it and the LKPY re-envisioning; and my students for their use and feedback in CS 538.
All errors of fact, judgement, or writing are my own.
\end{acks}

\bibliographystyle{ACM-Reference-Format}
\bibliography{lkpy}

\end{document}